\documentclass{elsart}
\usepackage{natbib}
\usepackage{graphicx}
\usepackage{amssymb}
\begin{document}
\begin{frontmatter}
\title{Self-Organized Criticality in a Bulk Driven One-Dimensional Deterministic
 System\thanksref{X}}
\author{Maria de Sousa Vieira\thanksref{permaddress}}
\address{International Center for Complex Systems, Departamento de Fisica, UFRN,59072 970 Natal, RN, Brazil.}
\thanks[X]{In honor of Constantino Tsallis in the celebration of his 60th birthday.}
\thanks[permaddress]{e-mail: mariav\_us@yahoo.com.} 
\begin{abstract}
We introduce a deterministic self-organized critical system that is
one dimensional and bulk driven. We find that there is no unique universality
class associated with the system. That is, the critical exponents change
as the parameters of the system are changed. This is in contrast with
the boundary driven version of the model [M. de Sousa Vieira, {\em Phys. Rev. E}
{\bf 61} (2000) 6056] in which the exponents are unique.
This model can be seen as a discretized version of 
the conservative limit of the Burridge-Knopoff model for earthquakes. 
\end{abstract}
\end{frontmatter}

Bak, Tang and Wiesenfeld\cite{soc} introduced the concept of Self-Organized 
Criticality (SOC) to explain the ubiquity of scaling invariance  in 
Nature. In systems that present SOC power-law distribution of event sizes,
limited only by the system size, 
is observed without fining tuning of parameters.

One of the most well known scaling invariant distributions in 
Nature is the Gutenberg-Richter law\cite{gutenberg}, which refers 
to a power-law distribution of earthquake sizes.  
Earthquake models have been   
studied in connection with SOC, one of them being the 
Burridge-Knopoff\cite{bk} 
model. That model consists of a linear chain of blocks
connected to each other via springs and each block is connected to 
an upper bar, which is pulled with constant velocity. The blocks 
are on a surface with friction. Events of several sizes are observed 
as the blocks are pulled. Carlson and Langer\cite{cl} showed that the distribution 
of events follows a power-law for small events. Larger events follow 
a different distribution. The critical event size that separates those 
two distributions does not depend on the system size. Since the power-law distribution 
is not limited by the system size, it implies that the Burridge-Knopoff model, in 
the nonconservative version studied by Carlson and Langer,  
does not present SOC. 
Another model for earthquakes, also 
introduced by Burridge and Knopoff\cite{bk}, in which only the first block 
of the chain is pulled  does present SOC. Such a model has been 
called the train model\cite{train}. 
A discretized version of the train model   
model governed by coupled map lattices (CML) was introduced in \cite{cmltrain}.
In a CML the space variables are continuous and the time is discrete.  
 
The aim of this paper is to show that  
a CML discretized version  of the Burridge-Knopoff model, the one in which 
all the blocks are connected to an upper bar, does present SOC in the 
conservative limit. Such a discretization  was performed by Nakanishi 
on the same model\cite{naka}. However, in his studies that 
system did not present SOC, even in 
the conservative limit, due to a different relaxation function he used. 
We also show that the critical exponents vary with the parameters of the 
model. This is in contrast with the train model, which has a unique universality
class. 
The universality class of the train model is the same as the one of the 
Oslo rice pile  model\cite{boettcher}. 
To our knowledge, this is the first SOC model to be introduced that 
is bulk driven, one-dimensional and deterministic. 
We believe that our model could also 
be applied for 
granular material in a quasi-one-dimensional  
rotating drum. 
The two-dimensional version of this model was studied in 
\cite{cml2train} and that model presents avalanche size distribution 
consistent with what is observed in sand experiments.   

In our discretized version of the Burridge-Knopoff model 
each element (i.e., blocks)  $i$ is characterized by a variable $f_i$, which we 
will call force, with  
$i=1,..., L$, and $L$ being  the number of elements in the system. 
The dynamical evolution of the system is determined by 
the following algorithm:

(1) Start the system by defining random initial values for the variables 
$f_i$,   
so the they are 
below a chosen, fixed,  threshold $f_{th}$.     

(2) Find the element in the system that has the largest $f$, 
denoted here by $f_{max}$.  
Then update all the elements according to    
$f_i \mapsto f_i+f_{th}-f_{max}$. 

(3) Check the forces in each  element. If an element $i$ has $f_i \ge f_{th}$, 
update  $f_i$ according to $f'_i=\phi(f_i-f_{th})$, where 
$\phi $ is a given nonlinear function.  
Increase the forces in its two nearest neighboring elements according 
to  $f'_{i \pm 1}=f_{i\pm 1}+\alpha \Delta f/2$, where 
$\Delta f=f_i-f_i'$ and $\alpha $ determines the level of conservation.   

(4) If $f'_i < f_{th}$ for all the elements, go to step (2) (the event 
has finished). Otherwise, go 
to step (3) (the event is still evolving).    

Without losing generality one can choose $f_{th}=1$. 
The functional form we use  for  $\phi$ is 
$\phi (x) = 1-d-a[x]$  
where $[x]$ denotes $x \ {\rm modulo} \ (1-d)/a$, 
that is, a sawtooth function. 
However, we have tested other periodic functions, such as a triangular wave,  
and found that the behavior we show here remains, that is,  
the results are robust, the essential ingredient being periodicity 
(not necessary a perfect one) 
for $\phi $.  The motivation for choosing a periodic function is due 
to the fact that we observed that in the Burridge-Knopoff model,  
modeled by ordinary differential equations (ODEs), the force in a 
block after a not so small  event  can have any value that is smaller than the  
maximum static friction force.  
Also, in the train model governed by CML only a periodic function 
reproduces what is seen in the system governed by ODEs, that is, SOC\cite{train,cmltrain}.
Nakanishi\cite{naka} used for $\phi $ a decreasing function, 
but that is unrealistic in our view for comparison with the system governed by ODEs. 

In our system, the parameter region in which SOC is found is $a>1$ and $d << 1$. For 
$a <  1$ one observes that the earthquakes merge with each other and 
the blocks never stop (for a discussion of the physical meaning of $a$ and $d$ and 
why $a=1$ represents an important 
boundary in such kind of models, see \cite{cmltrain,cml2train}). For the case $d = 1$ the system is not 
SOC and corresponds to the one-dimensional version of the OFC model\cite{ofc}. 
There is a transition at an intermediate values of $d$, which seems  to 
be around $d=0.7$, from SOC to non SOC behavior. It is beyond the scope of 
this paper to study such a transition. We simulate the system using 
open boundary conditions and parallel update. The number of avalanches 
we use is $10^6$ when $d \ge 0.1$ and $10^7$ when $d < 0.1$.     

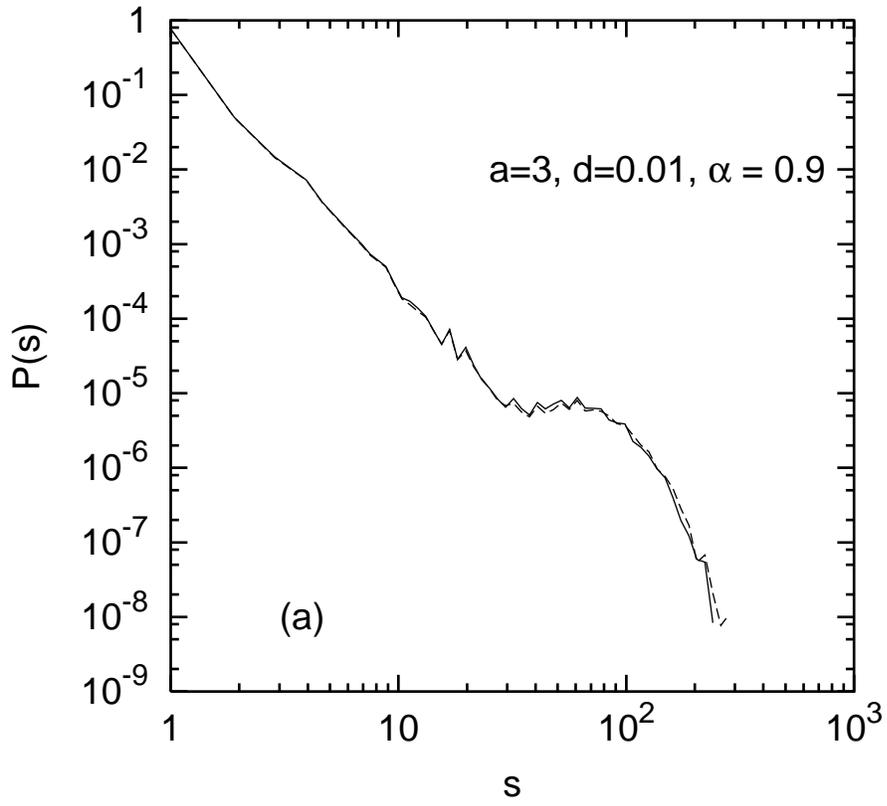
\begin{figure*}
  \vspace{3cm}
\caption{Distribution of event sizes for (a) $a=3$, $d=0.01$, $\alpha =0.9$ with $L=128$ (solid), $L = 256$ (dashed)  and (b) $a=4$, $d=0.1$ and $\alpha =1$.}
\label{f1}
\end{figure*}

Although the focus of this work is on the conservative case ($\alpha =1$) 
we first show an example of event size distribution in the case of non conservation.
We show in Fig.\ref{f1}(a) the frequency of events $P(s)$ as a function of the  
the event size $s$ when $\alpha = 0.9$. The qualitative behavior seen is the same observed in the model 
governed by ODEs\cite{cl}, that is, a  power-law
distribution for small events and  
a bumpy distribution for larger events. 
We have 
found that the events that belong to the power-law region are the ones 
that do not probe the large discontinuities of the relaxation function $\Phi $.
In other words, the only 
nonlinearity probed by a block whose force exceeds the threshold is a small discontinuity determined by $d$. 
We will be talking of this regime as the ``almost linear" one. We have 
found a similar situation in the system governed by ODEs. 
There, the bumpy part of the distribution has  the events that  
probe the nonlinear regime of the friction force.        

From now on we concentrate on the conservative case, i.e,  $\alpha =1$.
We show in Fig.\ref{f1}(b) a typical case 
of the avalanche distribution (solid line) in that regime. We have found that that distribution can be divided in two distinct distributions: one for the 
almost linear regime (short dashed line) and  one for the strong 
nonlinear regime (long dashed line). Power-law distribution limited only 
by the system size (i.e. SOC) occurs only in the strong nonlinear regime. 
We are going to concentrate our attention to that case  only. 

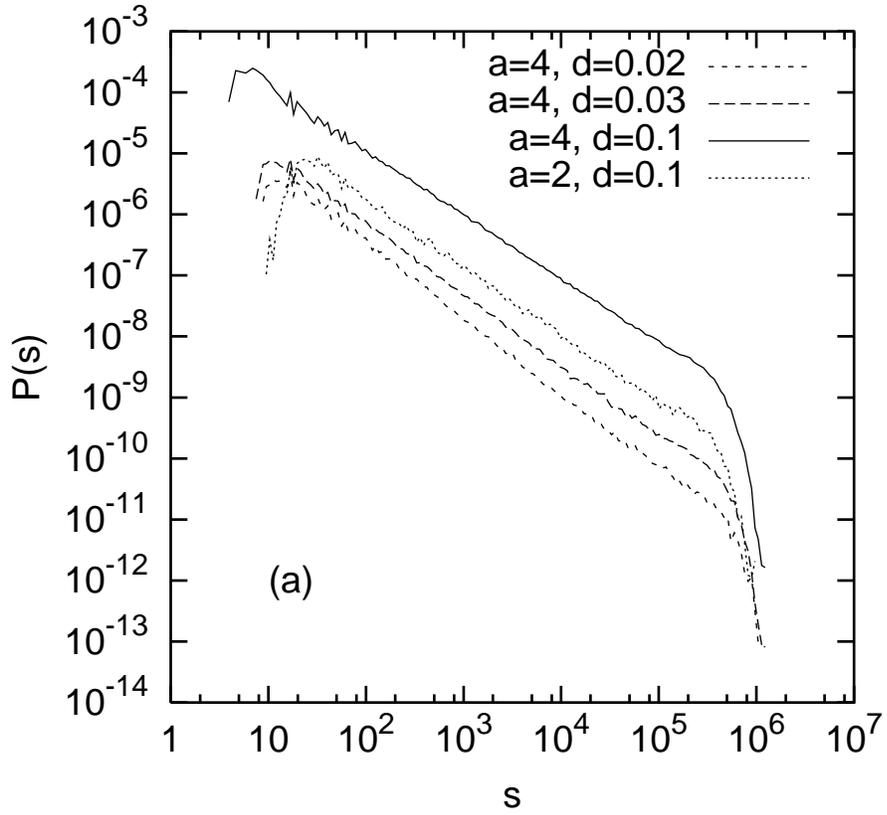
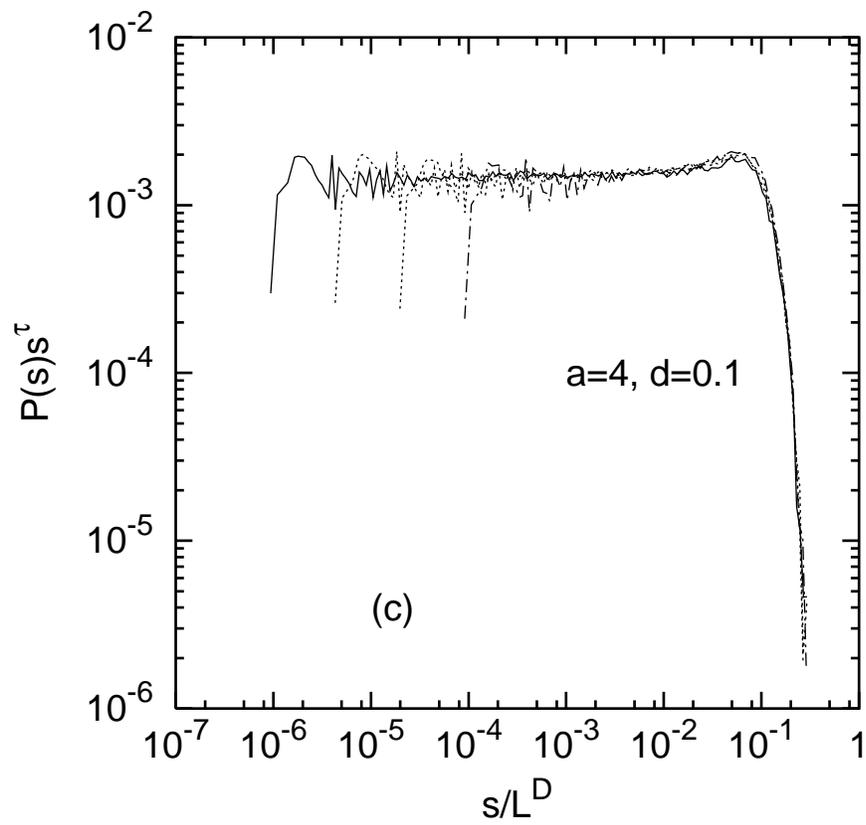
\begin{figure*}
  \vspace{3cm}
\caption{Distribution of event sizes for (a) varying $a$ and $d$ and keeping 
$L$ the same and (b) when $L=128, 256, 512, 1024$ with $a$ and $d$ kept constant. In (c) we show the finite-size scaling data collapse 
using $\tau =1.06$ and $D=2.20$.}
\label{f2}
\end{figure*}

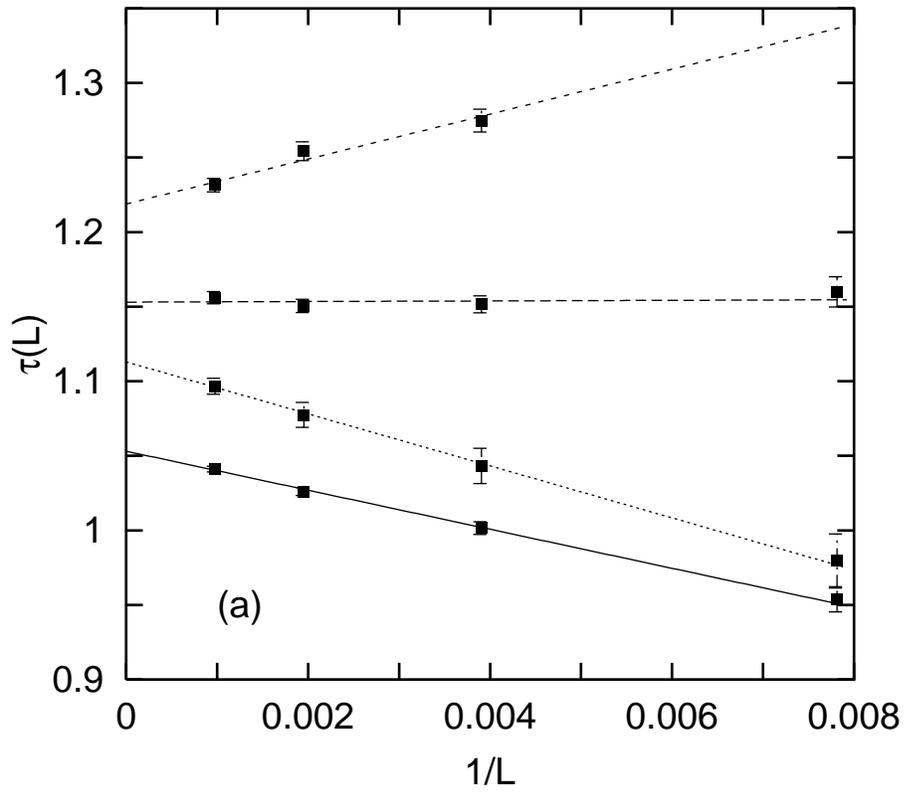
\begin{figure*}
  \vspace{3cm}
\caption{(a) Exponent $\tau $ as a function of $L$ for (from top to bottom) ($a=4$, $d=0.02$), ($a=4$, $d=0.03$), ($a=4$, $d=0.1$), ($a=2$, $d=0.1$) and (b) 
average value of the avalanches $<s>$ as a function of $L$.}
\label{f3}
\end{figure*}

We have plotted in Fig.\ref{f2}(a) the distribution of event sizes $P(s)$ for the 
events that are in the strong nonlinear regime for different values of 
$a$ and $d$. We see that $P(s) \sim s^{- \tau}$, which is characteristic 
of SOC systems. However, $\tau $ is not unique and vary with the system 
parameters. 
In Fig.\ref{f2}(b) we show $P(s)$ for different system sizes keeping $a$ and $d$ the same, and in Fig.\ref{f2}(c) we show the data collapse using the finite-size scaling ansatz for that set of parameters.  
We have noticed that the value of $\tau $ depends slightly on the system size, 
but converges to a given value as $L$ increases. This has been seen 
in the Zhang model\cite{lubeck}. The authors of \cite{lubeck} found 
that for the Zhang model the linear relation $\tau(L)=\tau _\infty - const/L$ is obeyed. By extrapolating $L \to \infty$ one obtains the value of $\tau $ for 
an infinite lattice. 
We have found the same in our model, and this is shown in Fig.\ref{f3}(a) 
for different parameter values.    

In many SOC models for sandpiles it has been observed that grains 
propagate as an unbiased random walker. This 
combined with a conservation law implies that $<s> \sim L^2$\cite{kadanoff}. 
In our model 
we have seen that this does not always occur. Instead, in the limit of large $L$, the relation $<s> \sim L^\mu $ with $\mu \le 2$ is observed. This means that the avalanche propagation 
in this model occurs subdifusively in some regions of the parameter space. The  
largest avalanche size scales as $s_{max} \sim L^D$. We have found that   
$D$ does not vary with the parameter values, having a value of $D \approx 2.20$. 

It is not difficult to show that $D$, 
$\mu $ and $\tau_\infty $ are related by $\mu = D(2 -\tau_\infty) $ (see for example \cite{sneppennaka} for a derivation when $\mu =2$). Consequently, if 
$D$ is constant and $\tau $ varies with the parameter values, this implies 
that $\mu $ cannot be same for all the parameter values. This is 
what Fig.~\ref{f3}(b) shows. We have estimated the asymptotic value of $\mu $ via 
$\mu = (\log <s_{L=1024}> - \log<s_{L=512}>)/(\log (1024)-\log(512)>)$, 
and we found the following values: $\mu = 1.67, 1.78, 2.05, 1.94$, for 
($a=4, d=0.02$), ($a=4, d=0.03$), ($a=4, d=0.1$), ($a=2, d=0.1$), respectively.  
Inputting the values of $\mu $ in $\mu = D(2 -\tau_\infty) $ we can find $\tau_\infty$ and the values of $\tau_\infty $ obtained in this way are in good 
agreement with what is shown in Fig.\ref{f3}(a).

I thank Maya Paczuski for enlightening discussions.

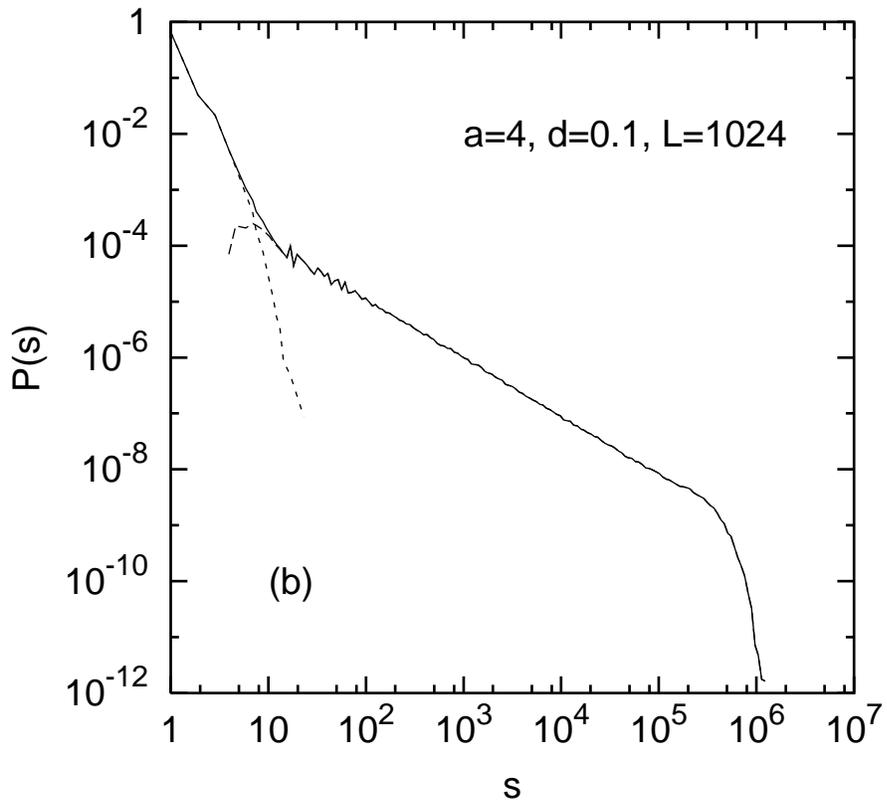
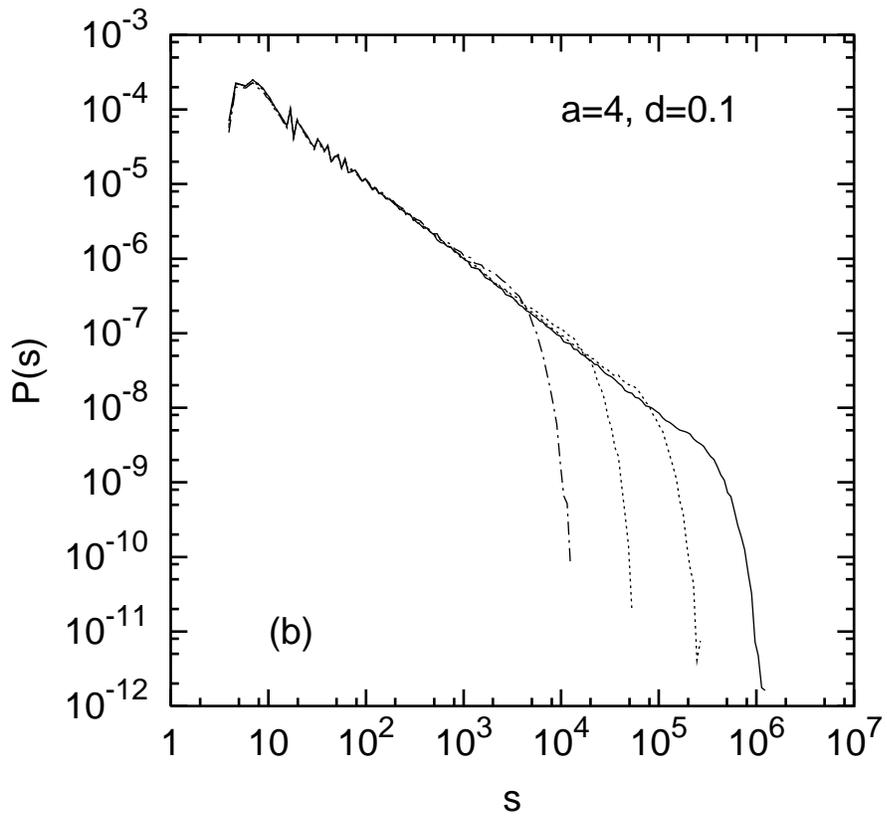
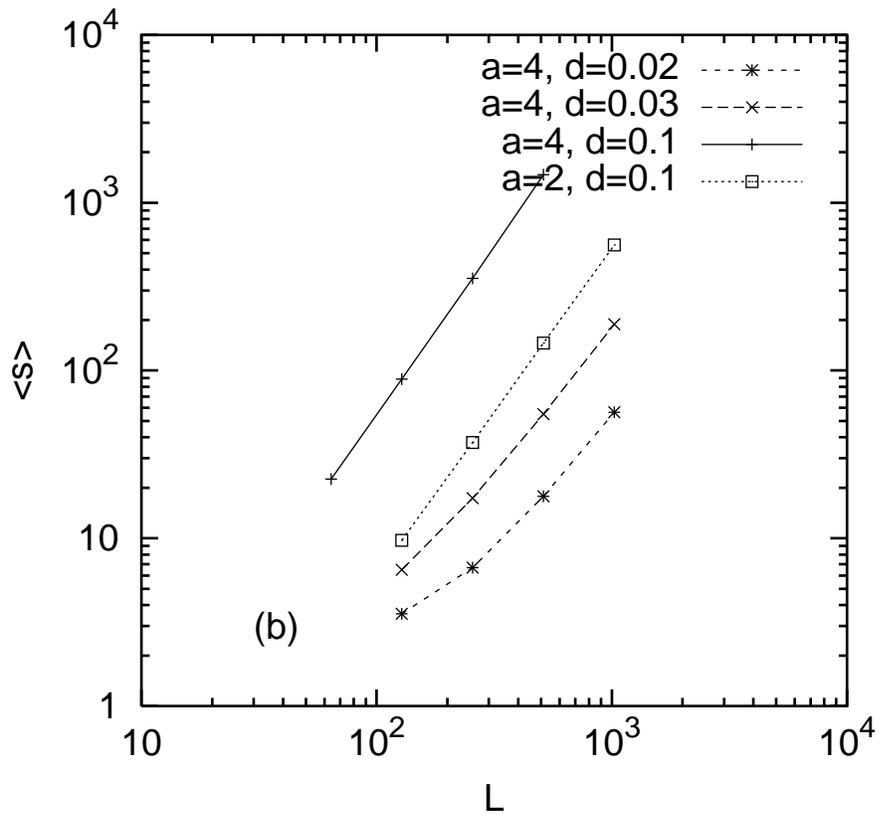
\end{document}